\documentclass{PoS}
\usepackage{amsmath}
\usepackage{slashed}
\usepackage{subfigure}

\newcommand{\formu}[1]{\(#1\)}
\newcommand{\ff}{f\mbox{}f}
\newcommand{\beq}{\begin{equation}}
\newcommand{\eeq}{\end{equation}}

\title{Renormalization constants for Wilson fermion lattice QCD with four dynamical flavours}

\ShortTitle{RCs -- four dynamical flavours}

\author{ETM Collaboration}
\author{Petros~Dimopoulos$^a$, Roberto~Frezzotti$^{b,c}$, Gregorio~Herdoiza$^{d}$,
        Karl~Jansen$^{d}$, Vittorio~Lubicz$^{e}$, \speaker{David~Palao}$^c$, Giancarlo~Rossi$^{b,c}$\\
        \llap{$^a$}Dipartimento di Fisica, Università di Roma ``La Sapienza''\\
        Piazzale A. Moro, I-00185 Rome, Italy\\
        \llap{$^b$}Dipartimento di Fisica, Università di Roma ``Tor Vergata''\\
        Via della Ricerca Scientifica 1, I-00133 Rome, Italy\\
        \llap{$^c$}INFN Sezione di ``Roma Tor Vergata''\\
        c/o Dipartimento di Fisica, Università di Roma ``Tor Vergata''\\
        Via della Ricerca Scientifica 1, I-00133 Rome, Italy\\
        \llap{$^d$}DESY, Platanenallee 6, D-15738 Zeuthen, Germany\\
        \llap{$^e$}Dipartimento di Fisica, Università di Roma Tre and INFN\\
        Via della Vasca Navale 84, I-00146 Rome, Italy\\
        E-mail: \email{david.palao@roma2.infn.it}}

\abstract{We report on an ongoing non-perturbative computation of RI-MOM scheme renormalization constants for 
the lattice action with four dynamical flavours currently in use by ETMC.
For this goal dedicated simulations with four degenerate sea quark flavours
are performed at several values of the standard and twisted quark mass parameters. We discuss a
method for removing possible O(a) artifacts at all momenta and extrapolating renormalization 
constant estimators to the chiral limit. We give preliminary results at one lattice spacing. }

\FullConference{The XXVIII International Symposium on Lattice Field Theory, Lattice2010\\
		June 14-19, 2010\\
		Villasimius, Italy}

\begin{document}

\section{Introduction}
Simulations including two degenerate light flavours and
   a non-degenerate doublet of quarks are currently being
   performed by the European Twisted Mass (ETM) Collaboration. The
   inclusion of $n_f=2+1+1$ flavours is a necessary step to move towards a
   realistic situation. 
   Fermions are described by the maximally twisted mass
   lattice QCD (MtmLQCD) action \cite{tmW-2003}
   and gluons by the Iwasaki action \cite{Iwasaki:1985we}.
   While the first physical
   results are very encouraging \cite{MtmNf4-2010},
 dedicated
   simulations are required to perform the non-perturbative
   re\-nor\-ma\-li\-za\-tion of operators in a mass-independent scheme,
   where renormalization constants (RCs) are defined at zero quark mass.
   In the study of $n_f=2+1+1$ QCD 
   ETMC is adopting the
   RI-MOM scheme~\cite{Martinelli:1994ty}. 
   The RCs are evaluated by extrapolating to the chiral limit
   the RC estimators computed in the theory with $n_f=4$ mass 
   degenerate quarks for a range of mass values~\footnote{For
   Monte Carlo simulations we used a highly optimized implementation of a 
   HMC-like algorithm~\cite{Jansen:2009xp}.}.  
   Here we report on the progress we made in this project.

\subsection{Action and quark mass parameters} 
For the present study we consider the lattice action
\beq
S_L=S^{\rm YM}_{Iwa}+ a^4 \sum_{x}\sum_{f=1}^4 \bar\chi_f \left[ \gamma\cdot\widetilde{\nabla}
- \tfrac{a}{2} \nabla^*\nabla + m_{0,f} + i\gamma_5 r_f \mu_f \right]\chi_f(x)
\eeq
where \formu{\chi_f} is a one-flavour quark field in the so-called \textit{twisted basis} and
in this work $r_f$ is set to either $1$ or $-1$.
Passing from the twisted to the physical quark basis\footnote{The 
relation between twisted (\formu{\chi_f} fields) and physical (\formu{q_f} fields) quark basis is
\beq
\chi_f\to q_f = \exp [\tfrac{i}{2}(\tfrac{\pi}{2} -\theta_{0,f}) \gamma_5 r_f] \chi_f\,,\quad
\bar{\chi}_f\to \bar{q}_f = \bar{\chi}_f\exp [\tfrac{i}{2}(\tfrac{\pi}{2} -\theta_{0f}) 
\gamma_5 r_f]
\eeq
}
\beq
S_L=S^{\rm YM}_{Iwa} + a^4 \sum_{x}\sum_{f=1}^4 \bar{q}_f \left[ \gamma\cdot\widetilde{\nabla}
-i\gamma_5 r_f e^{i\gamma_5 r_f\theta_{0,f}} (- \tfrac{a}{2} \nabla^*\nabla + m_{\rm cr}) 
+ M_{0,f} \right] q_f(x)  \, .
\eeq
The bare mass parameters can be rewritten as
\beq
M_{0,f} = \sqrt{ (m_{0,f} - m_{\rm cr})^2 + \mu_f^2 }\,,\quad
\sin \theta_{0,f} = \frac{m_{0,f} - m_{\rm cr}}{M_{0,f}}\,, \quad
\cos \theta_{0,f} = \frac{\mu_f}{M_{0,f}}\,.
\eeq
Their renormalized counterparts read
\formu{M_f = Z_P\hat{M}_f = \sqrt{ Z_A^2 m_{\rm PCAC}^2 + \mu_f^2 }} and 
\formu{\tan \theta_f = \frac{ Z_A m_{\rm PCAC} }{\mu_f}}. The parametrization 
in terms of \formu{M} and \formu{\theta} is 
convenient because the leading term of the Symanzik local e\ff ective Lagrangian involves only 
\formu{M}, not \formu{\theta}. As we will see later (see the end of section \ref{strategy}), 
this remark is at the basis of our method to obtain 
\(O(a)\)-improved RC-estimators at all scales even  \textit{out of maximal twist}.
Since, for practical reasons, we work in a partially quenched setup with all four flavours
having equal mass parameters, we will have to consider in our analysis four quark mass 
parameters: \formu{M_{\rm sea},\theta_{\rm sea}, M_{\rm val},\theta_{\rm val}}.

\subsection{RI'-MOM scheme and our setup}
The focus of the present study is on flavour non-singlet quark bilinear operators,  
\formu{O_\Gamma=\bar{\chi}_f \Gamma \chi_{f'}} (or \formu{\bar{\chi}_{f'}\Gamma \chi_f}), 
with \formu{\Gamma\;=\;S, P, V, A, T}, which are written
in terms of $\chi$ and $\bar\chi$ quark fields (i.e. in the 
standard quark basis for untwisted Wilson fermions). 
RCs are named after the expression of the operators in this basis so as
to match the usual notation in the literature about Wilson fermions.\\ 
As convenient in lattice studies, we adopt the
RI'-MOM scheme~\cite{Franco:1998bm,Martinelli:1994ty}, which is defined as follows. 
A first condition fixes the quark field renormalization, namely 
\beq
Z_q^{-1}\frac{-i}{12N(p)}\sum_\rho\mbox{}^\prime\left[ 
\frac{\mathrm{Tr}(\gamma_\rho S_f(p)^{-1})}{\tilde{p}_\rho}\right]_{\tilde{p}^2=\mu^2} 
\,=\,1 \, , \qquad 
\mathrm{any }f  \, ,
\eeq
where \formu{\tilde{p}^2=\sum_\mu\tilde{p}_\mu^2\,,\ \tilde{p}_\mu\equiv\tfrac{1}{a}\sin ap_\mu}. The sum \formu{\sum_\rho^\prime} only runs over the Lorentz indices for which
\formu{p_\rho} is di\ff erent from zero and \formu{N(p)=\sum_\rho^\prime 1}. The 
renormalization condition for the operators \formu{O_\Gamma} reads
\beq
Z_q^{-1}Z^{(ff^\prime)}_\Gamma\mathrm{Tr}\left[
\Lambda^{(ff^\prime)}_\Gamma(\tilde{p},\tilde{p})P_\Gamma\right]_{\tilde{p}^2=\mu^2} 
\,=\,1  \, ,  \qquad f\neq f^{\prime} \, .
\label{ZGam} \eeq
Above $S_f(p)\,=\,a^4\sum_x\,e^{-ipx}\left\langle \chi_f(x)\bar{\chi_f}(0)\right\rangle$
is the $\chi_f$ field propagator in momentum space,
while
\beq
\Lambda^{(ff^\prime)}_\Gamma(p,p)\,=\,
S_f^{-1}(p)G_\Gamma^{(ff^\prime)}(p,p)S_{f^\prime}^{-1}(p)
\eeq
denotes the quark bilinear vertex that is obtained by ``amputating'' the Green function
\beq
G^{(ff^\prime)}_\Gamma(p,p)\,=\,a^8\sum_{x,y}\,e^{-ip(x-y)}
\left\langle \chi_f(x)(\bar{\chi}_f\Gamma\chi_{f^\prime})(0)\bar{\chi}_{f^\prime}(y)\right\rangle 
\qquad\Gamma\;=\;S, P, V, A, T \, .
\eeq
Barring cutoff effects,
RCs are independent of \formu{sign(r_f)}. For practical reasons
here we limit ourselves to \formu{r_{f^\prime}=-r_f} in evaluating \formu{Z_\Gamma}, see
eq.~(\ref{ZGam}). 
\section{Strategy for RCs in the \formu{N_f=4} theory}\label{strategy}
In order to extract useful information from simulations performed with twisted 
mass Wilson fermions one 
must know the twist angle, \formu{\omega=\tfrac{\pi}{2}-\theta}, 
with good precision. The level of precision requested for $\omega$  depends  
on the observable of interest. In our case, after
an exploratory study on a few \formu{16^3\times 32} lattices \cite{Deuzeman:2009zz}, and some 
tests near maximal twist on a \formu{24^3\times48} lattice 
we have chosen to work \textit{out of maximal twist}.

\begin{figure}[!ht]\label{closetometa}
\centering
\subfigure[]{
\label{pcacVSm0}\includegraphics[scale=0.35]{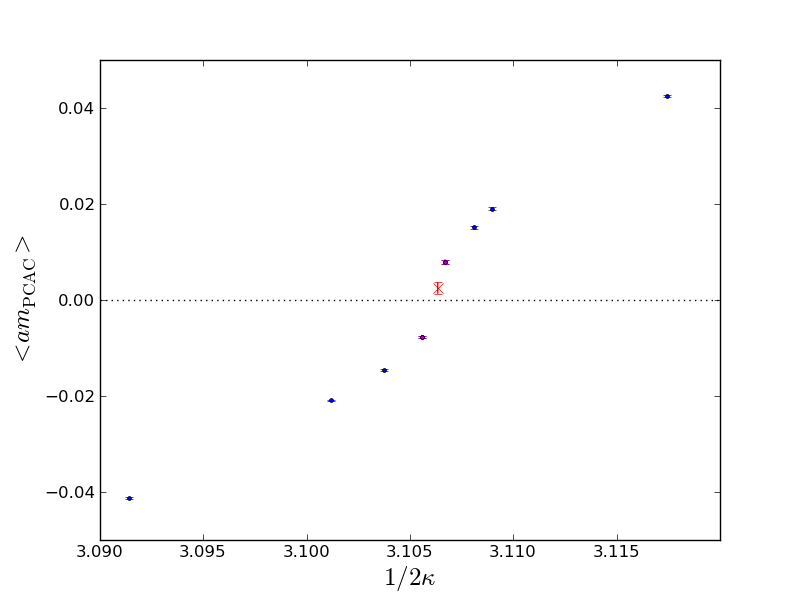}
}
\subfigure[]{
\label{pcacVSt}\includegraphics[scale=0.35]{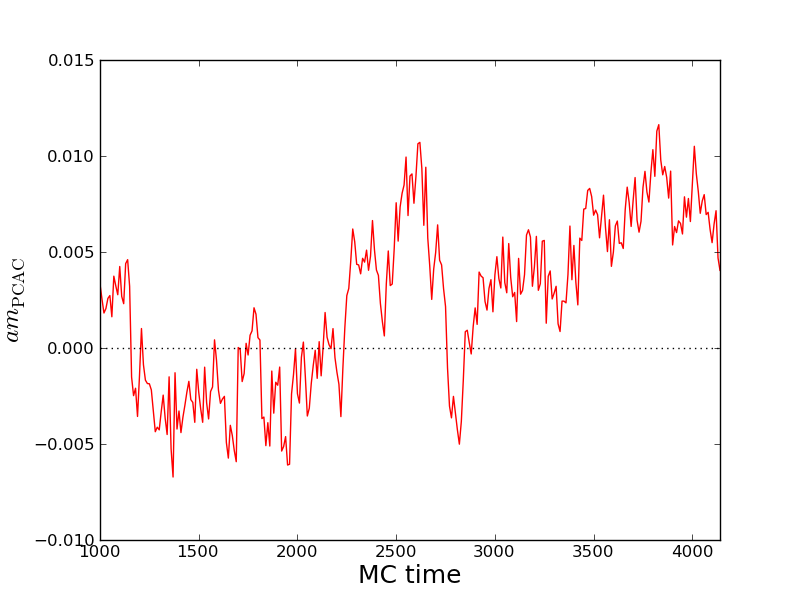}
}\caption{(a) \formu{am_\mathrm{PCAC}} versus \formu{1/2\kappa} on lattices
\formu{24^3\times 48} at $\beta=1.95$; 
(b) Monte Carlo history of $am_{\rm PCAC}$ for the most critical case
(corresponding to the red cross point) in panel (a).}
\end{figure}

Figure \ref{closetometa} illustrates the di\ff iculties of tuning to maximal twist,
i.e.\ setting $m_{\rm PCAC}$ to zero, in the simulation setup for RC computations,
at least if the lattice spacing is not very fine. 
Specifically, the slope of $m_{\rm PCAC}$ vs $1/(2\kappa)$ in figure \ref{pcacVSm0} 
suggests that near $m_{\rm PCAC}=0$ simulations  
are in a region with a sharpe change of the slope for \formu{m_\mathrm{PCAC}}
where it is di\ff icult to extract 
useful information. On the other hand
figure \ref{pcacVSt} gives a more quantitative view of this problem showing results from
one simulation close to the critical point (the point closest to \formu{m_{\mathrm{PCAC}}=0}
in figure \ref{pcacVSm0}). It appears that 
due to the long fluctuations a precise measurement of the PCAC mass will require for this case
a very large number of Monte Carlo trajectories. In fact, we have observed a similar feature
for all the ensembles with \formu{|am_\mathrm{PCAC}|\lesssim 0.01} 
at both \formu{\beta=1.95} and \formu{\beta=1.90} (i.e.\ for $a \geq 0.08$~fm).

In summary, working at maximal twist for the chosen range of twisted masses 
(see table~1) would imply a considerable fine tuning work
owing to the di\ff iculties in determining $am_\mathrm{PCAC}$ near $am_\mathrm{PCAC}=0$.
To alleviate the problem one would need to increase the value
of the twisted mass, $\mu_f$, and thus $M_f$. Instead, working away from
\textit{maximal twist}, one can avoid the metastable region of parameter space
and measure the twist angle with good precision. This comes at the price
of a moderate increase of the quark mass $M_f$ and of a slightly more involved
analysis. In our RC-estimators cutoff effects linear in $a$ are expected to be 
small and 
can anyway be removed with 
controlled precision by averaging the results 
obtained for a given $M_f$ at opposite values of $\theta_f$.

In fact, from the symmetry of the lattice action \formu{S_L} under
\formu{{\cal P} \times (\theta_0 \to -\theta_0) \times 
{\cal D}_d \times (M_0 \to - M_0)} \cite{tmW-2003,Frezzotti:2005gi,RCnf4}
it follows that the \formu{O(a^{2k+1})} artifacts occurring in the vacuum expectation
values of 
(multi)local operators \formu{O}
that are invariant under \formu{{\cal P } \times (\theta_0 \to -\theta_0)} are 
quantities that change sign upon changing the sign of \formu{\theta_0} (or \formu{\theta}).
Hence \formu{O(a^{2k+1})} cuto\ff\ e\ff ects vanish in \(\theta\)-averages: \formu{~\frac{1}{2} \Big[
\langle O \rangle|_{\hat{M},\theta} + \langle O \rangle|_{\hat{M},-\theta} \Big]}.
The same is true for operator form factors invariant 
under \formu{{\cal P } \times (\theta_0 \to -\theta_0)} and, in particular, 
for our RC-estimators at all values of \formu{M_f} 
and \formu{\tilde{p}^2}.


\section{Current analysis and preliminary results}

\begin{table}[!ht]\label{runs}
\begin{center}
\begin{tabular}{ccccccc}
ensemble & \formu{a\mu_\mathrm{sea}} & \formu{am_\mathrm{PCAC}^\mathrm{sea}} & \formu{aM_0^\mathrm{sea}} 
& \formu{\theta^\mathrm{sea}} & \formu{a\mu^\mathrm{val}} & \formu{am_\mathrm{PCAC}^\mathrm{val}} \\

\hline
\texttt{1m} & 0.0085 & -0.04125(13) & 0.03288(10) & -1.3093(8) & [0.0085,\ldots, 0.0298] & -0.0216(2)\\
\texttt{1p} & 0.0085 & +0.04249(13) & 0.03380(10) & 1.3166(7) & [0.0085,\ldots, 0.0298] & +0.01947(19)\\
\hline
\texttt{3m} & 0.0180 & -0.0160(2) & 0.02182(9) & -0.601(6) & [0.0060,\ldots, 0.0298] & -0.0160(2)\\
\texttt{3p} & 0.0180 & +0.0163(2) & 0.02195(9) & 0.610(6) & [0.0060,\ldots, 0.0298] & +0.0162(2)\\
\hline
\texttt{2m} & 0.0085 & -0.02091(16) & 0.01821(11) & -1.085(3) & [0.0085,\ldots, 0.0298] & -0.0213(2)\\
\texttt{2p} & 0.0085 & +0.0191(2) & 0.01696(16) & 1.046(6) & [0.0085,\ldots, 0.0298] & +0.01909(18)\\
\hline
\texttt{4m} & 0.0085 & -0.01459(13) & 0.01409(8) & -0.923(4) & [0.0060,\ldots, 0.0298] & -0.01459(13)\\
\texttt{4p} & 0.0085 & +0.0151(2) & 0.01441(14) & 0.940(7) & [0.0060,\ldots, 0.0298] & +0.0151(2)\\
\hline
\end{tabular}
\end{center}\caption{Mass parameters of the ensembles analysed for this contribution.
From the formulae in sect. 1.1 it follows that in the valence sector we have 
\formu{~~0.013 \lesssim aM^{\rm val} \lesssim 0.033}
and \formu{0.4 \lesssim |\theta_{\rm val}| \lesssim 1.2~~}
(\formu{\theta_{\rm val}/m_{\rm PCAC}^{\rm val} > 0}). }
\end{table}

Here we detail the analysis procedure we followed in order to obtain
{\em very preliminary} results on the RCs of interest. Indeed, at this
stage our main goal was checking the feasibility of the project.
In particular, the analysis procedure is not yet the optimal one, for instance
concerning the order of the various steps, and some refinements, such as the 
subtraction of the known cutoff effects at O($a^2g^2$)~\cite{Constantinou:2009tr},
are still omitted. While these improvements will be included in the
final analysis, the present work shows that the strategy advocated in 
section~2  allows to extract the RCs of the quark field and quark bilinear
operators with a $\sim 1\%$ level precision by means of stable simulations
at a lattice spacing ($a\sim 0.08$~fm) which is among the coarsest ones explored
in the study of $n_f=2+1+1$ QCD by ETMC. 

In practice, for a sequence of $M_\mathrm{sea}$-values,
we produced for each $M_\mathrm{sea}$ two ensembles 
with opposite values of \formu{\theta_\mathrm{sea}}.
We label them as \texttt{E}p/m,  where \texttt{E}\formu{\,=1,2\ldots} and 
p/m refers to sign(\formu{\theta_\mathrm{sea}}).
On each ensemble \texttt{E}p/m, with \formu{(M_\mathrm{sea}^{\mathtt{E}\mathrm{p/m}},
\theta_\mathrm{sea}^{\mathtt{E}\mathrm{p/m}})}
we compute the RC-estimators for several values of the
valence mass parameters \formu{(M_{\rm val},\theta_{\rm val})}
and \formu{\tilde{p}^2} (all corresponding to ``democratic'' momenta $p$,
in the sense specified in \cite{Constantinou:2010gr}), as 
summarized in table~1. 
Then we proceed in various steps as follows.

\begin{figure}
\centering
\subfigure[]{
\label{sub_gold}\includegraphics[scale=0.25]{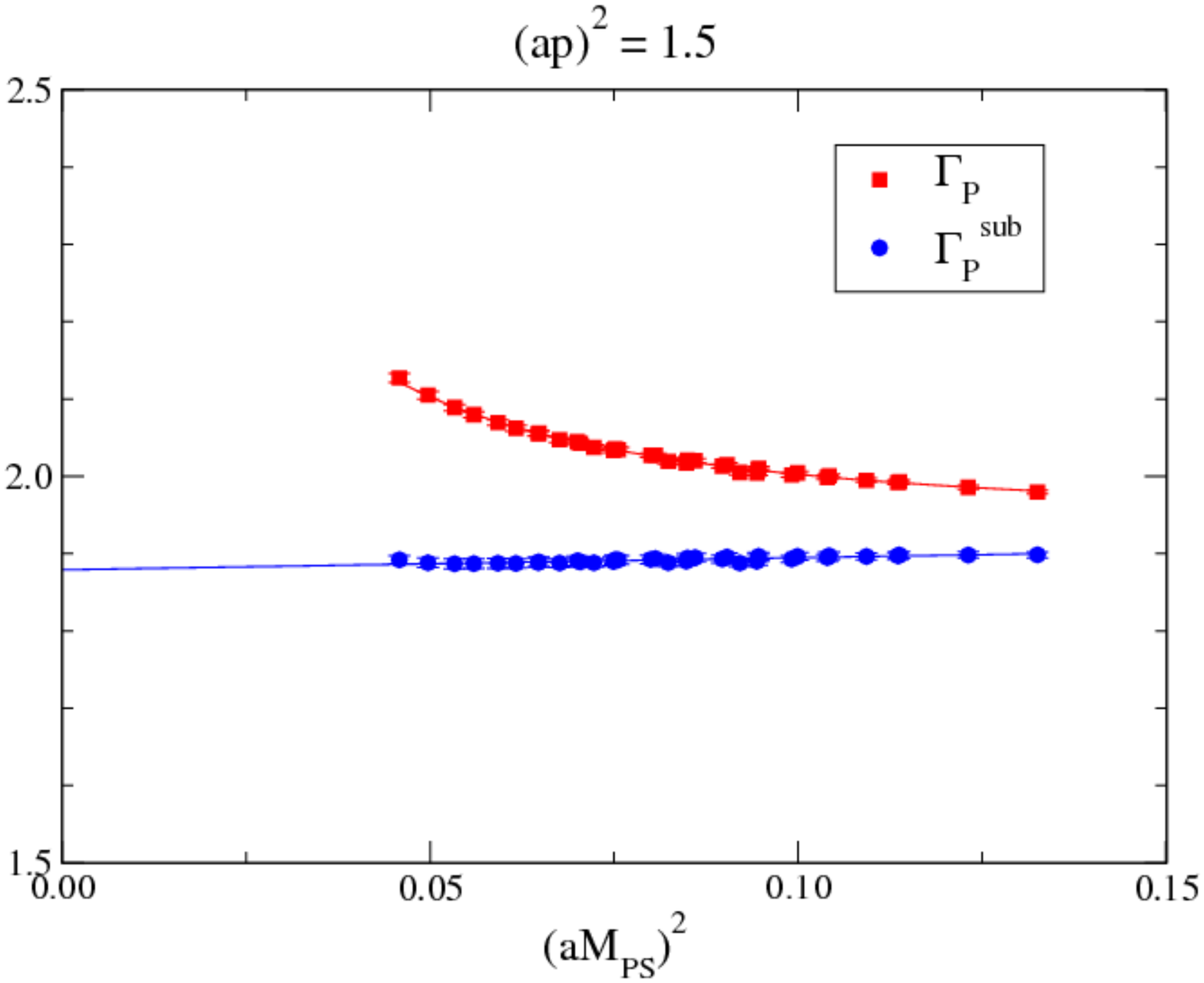}
}
\subfigure[]{\label{val_4m}\includegraphics[scale=0.54]{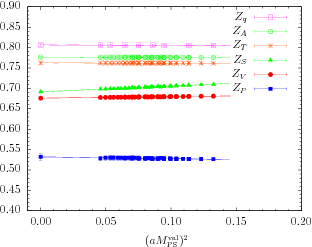}
}
\label{valence}\caption{For the example of ensemble 4m, $(a\tilde{p})^2=1.5$: 
(a) subtraction of Goldstone pole
contribution and valence chiral extrapolation in 
\formu{\Gamma_P=\mathrm{Tr}\left[\Lambda_P P_P\right]}; 
(b) overview of the valence chiral extrapolation for all RCs.}
\end{figure}

\begin{description}
\item[Valence chiral limit.] A fit of RC-estimators linear in 
\formu{(M^{PS}_{\rm val})^2} turns out to be numerically 
adequate (see fig.~\ref{val_4m}). For $\Gamma = P$ 
(see fig.~\ref{sub_gold}) or, due to O($a^2$) terms, $\Gamma=S$,
we have also kept into account the contribution $\propto 
(M^{PS}_{\rm val})^{-2}$ coming from the Goldstone boson pole.
\item[\formu{O(a^2\tilde{p}^2)} discretization errors.] We applied two
different methods, following \cite{Constantinou:2010gr}. 
In the first method (``M1''),
after bringing, via the known~\cite{PTevol-2003} perturbative evolution
the RC-estimators to a common renormalization scale 
(\formu{\tilde{p}^2_\mathrm{M1}=1/a^2}), we remove the remaining
\formu{O(a^2\tilde{p}^2)} discretization errors by a linear fit in 
$\tilde{p}^2$. 
The second method (``M2'') 
consists in simply taking 
the value of the RCs estimators at a high momentum point fixed in physical units. We chose 
\formu{\tilde{p}^2_\mathrm{M2}\,=\,12.2~\mathrm{GeV}^2}. The two approaches 
yield RC results di\ff ering only by cuto\ff\ e\ff ects.\\ 

\begin{figure}
\centering
\subfigure[]{
\includegraphics[scale=0.55]{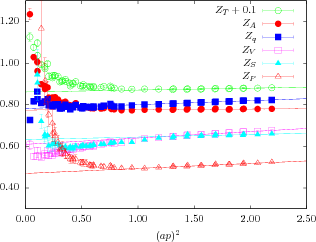}}
\hspace{0.5cm}
\subfigure[]{
\includegraphics[scale=0.55]{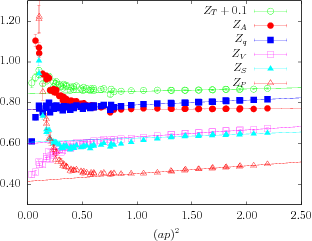}}
\caption{
Residual $\tilde{p}^2$-dependence of RC-estimators at 
scale $1/a$ and RC values from method M1 for the cases
of ensemble 1m (panel (a)) and 2p (panel (b)).
}
\end{figure}

\item[Removal of \formu{O(a)} artifacts.] It is achieved by  
\(\theta\)-average (see section~\ref{strategy}) of the RCs estimators,
\beq
Z_\Gamma (M_\mathrm{sea}^\mathtt{E}, 
|\theta_\mathrm{sea}^\mathtt{E}|)\,=\,
\frac{1}{2} \Big[
\langle Z_\Gamma (\hat{M}_\mathrm{sea}^{\mathtt{E}\mathrm{p}},
\theta_\mathrm{sea}^{\mathtt{E}\mathrm{p}}; 
\theta_\mathrm{val;eff}^{\mathrm{E}\mathrm{p}} )
\rangle + \langle Z_\Gamma (\hat{M}_\mathrm{sea}^{\mathtt{E}\mathrm{m}},
\theta_\mathrm{sea}^{\mathtt{E}\mathrm{m}}; 
\theta_\mathrm{val;eff}^{\mathrm{E}\mathrm{m}} )
 \rangle \Big]
\eeq
where $\theta_\mathrm{val;eff}^{\mathrm{E}\mathrm{p(m)}}$ parameterizes the 
dominating O($a$) effects in RC-estimators that (in the present analysis)  
arise from employing $M_{\rm val;Ep(m)}^{\rm PS}$ in the valence chiral
extrapolation.  

\item[Sea chiral limit.] The quantities $Z_\Gamma (M_\mathrm{sea}^\mathtt{E}, 
|\theta_\mathrm{sea}^\mathtt{E}|)$ are extrapolated to $M_{\rm sea}=0$ by
using the fit Ansatz 
\beq
Z_\Gamma(M_\mathrm{sea},\theta_\mathrm{sea})\,=\,Z_\Gamma\;+\;A\,M^2_\mathrm{sea}\;
+\;B\,M^2_\mathrm{sea}\,\cos(2\theta_\mathrm{sea}) \, .
\eeq
This Ansatz can be justified by an analysis \`a la Symanzik of the lattice
artifacts in $Z_\Gamma(M_\mathrm{sea},\theta_\mathrm{sea})$ up to O($M_{\rm sea}^2$)
and neglecting chiral spontaneous symmetry breaking effects~\cite{RCnf4}.
\end{description}
\begin{figure}
\centering
\subfigure[]{
\includegraphics[scale=0.35]{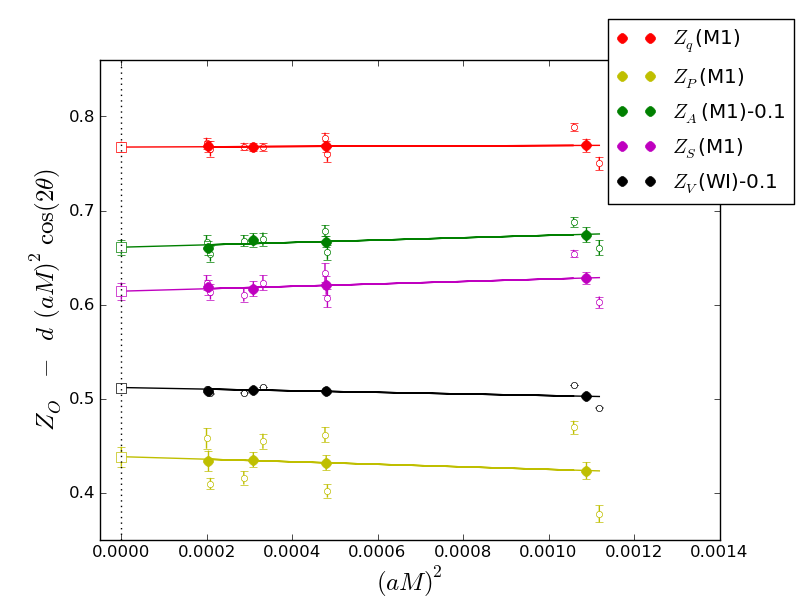}}
\subfigure[]{
\includegraphics[scale=0.35]{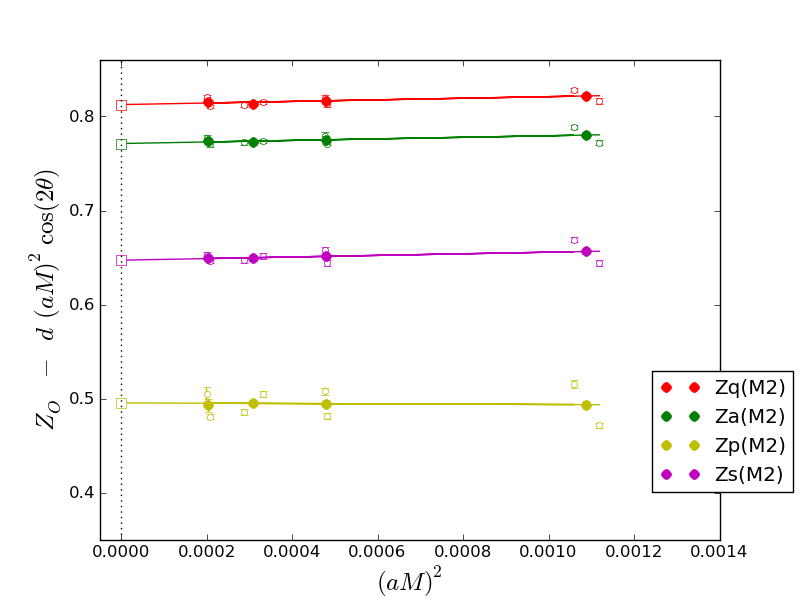}}
\caption{$M^{\rm sea}$-dependence before (empty symbols) and after 
(full symbols) $\theta$-average of RC-estimators for a few operators
as well as their chiral limit value. 
The left (right) panel corresponds to M1 (M2) results. In (a)  
we also show $Z_V({\rm WI})$, which is obtained by 
exploiting an exact lattice Ward-Takahashi identity (WI).} 
\end{figure}
The first, very preliminary results of this analysis are summarized in table~2.

\begin{table}[!ht]
\begin{center}
\begin{tabular}{cccccccc}
\hline \hline
Method & $Z_A$ & $Z_V$ & $Z_P(1/a)$ & $Z_S(1/a)$ & $Z_P/Z_S$ & $Z_T(1/a)$ & $Z_q(1/a)$ \\
\hline
&&&&&&& \\
M1 & 0.761(08) & 0.630(05) &  0.438(08) & 0.614(09) & 0.716(21) & 0.753(07) & 0.767(06) \\
M2 & 0.771(03) & 0.674(03) & 0.496(04) & 0.647(03) & 0.767(08) & 0.768(03) & 0.813(02) \\
&&&&&&& \\
\hline
\end{tabular}
\end{center}
\label{results}\caption{Preliminary RC results  
at \formu{\beta=1.95} from the analysis of section~3.
We also get $Z_V({\rm WI}) = 0.612(1)$.} 
\end{table}
\section{Conclusions and outlook}\label{conclusions}
We have described our strategy to compute O($a$) improved operator RCs
for the \formu{N_f=4} lattice action currently used by ETMC. 
We have shown that the method advocated in this work provides very 
encouraging results at one lattice spacing ($a \sim 0.08$~fm) that is among 
the coarsest simulated in the study of QCD with $n_f=2+1+1$ dynamical flavours.
In particular, the observed dependences of RCs on valence and sea quark masses are
mild and quite in line with our experience~\cite{Constantinou:2010gr} in 
\formu{n_f=2} QCD. Besides the technical improvements mentioned in section~3,
we plan to possibly add few more
ensembles at $a \sim 0.08$~fm ($\beta=1.95$) and to
extend our work to other lattice spacings.

We thank \href{http://www.idris.fr/}{IDRIS} and
\href{http://apegate.roma1.infn.it/APE/}{INFN/apeNEXT} 
for giving us CPU time necessary for this study.


\begin{thebibliography}{99}

\bibitem{tmW-2003}
  R.~Frezzotti and G.~C.~Rossi,
  JHEP {\bf 0408} (2004) 007
  [arXiv:hep-lat/0306014]
  and  
  Nucl.\ Phys.\ Proc.\ Suppl.\  {\bf 128} (2004) 193
  [arXiv:hep-lat/0311008].


%

\bibitem{Iwasaki:1985we}
  Y.~Iwasaki,
  Nucl.\ Phys.\  B {\bf 258} (1985) 141.

\bibitem{MtmNf4-2010}
  R.~Baron {\it et al.} [ETM Collaboration],
  JHEP {\bf 1006} (2010) 111
  [arXiv:1004.5284 [hep-lat]]
  and
  arXiv:1009.2074 [hep-lat];
  F.~Farchioni {\it et al.} [ETM Collaboration],
  arXiv:1012.0200 [hep-lat];
  G.~Herdoiza, PoS (Lattice 2010) 010

%
%
%
 

\bibitem{Martinelli:1994ty}
  G.~Martinelli, C.~Pittori, C.~T.~Sachrajda, M.~Testa and A.~Vladikas,
  Nucl.\ Phys.\  B {\bf 445} (1995) 81
  [arXiv:hep-lat/9411010].


\bibitem{Jansen:2009xp}
  K.~Jansen and C.~Urbach,
  Comput.\ Phys.\ Commun.\  {\bf 180} (2009) 2717
  [arXiv:0905.3331 [hep-lat]].

\bibitem{Franco:1998bm}
 E.~Franco and V.~Lubicz,
 Nucl.\ Phys.\  B {\bf 531} (1998) 641
 [arXiv:hep-ph/9803491].

\bibitem{Deuzeman:2009zz}
  A.~Deuzeman {\it et al.} [ETM Collaboration]
  PoS {\bf LAT2009} (2009) 037.

\bibitem{Frezzotti:2005gi}
  R.~Frezzotti, G.~Martinelli, M.~Papinutto and G.~C.~Rossi,
  JHEP {\bf 0604} (2006) 038
  [arXiv:hep-lat/0503034].

\bibitem{Constantinou:2009tr}
  M.~Constantinou, V.~Lubicz, H.~Panagopoulos and F.~Stylianou,
  JHEP {\bf 0910} (2009) 064
  [arXiv:0907.0381 [hep-lat]].

\bibitem{Constantinou:2010gr}
  M.~Constantinou {\it et al.}  [ETM Collaboration],
  JHEP {\bf 1008} (2010) 068
  [arXiv:1004.1115 [hep-lat]].

\bibitem{PTevol-2003}
  J.~A.~Gracey,
  Nucl.\ Phys.\  B {\bf 662} (2003) 247
  [arXiv:hep-ph/0304113];
  K.~G.~Chetyrkin and A.~Retey,
  Nucl.\ Phys.\  B {\bf 583} (2000) 3
  [arXiv:hep-ph/9910332].

%
%

\bibitem{RCnf4}
  ETM Collaboration, in preparation
\end{thebibliography}
\end{document}